\documentclass[10pt]{article}
\usepackage{geometry}                % See geometry.pdf to learn the layout options. There are lots.
\usepackage{graphicx}
\usepackage{amsmath}
\usepackage{amssymb}
\usepackage{latexsym}
\usepackage{epstopdf}
\newcommand{\sns}{\sum\limits_{S \sim S^\prime}}

\newcommand{\tausp}{\psi_{S^\prime} (t)}
\newcommand{\taus}{\psi_{S} (t)}
\newcommand{\tauso}{\psi_{S} (0)}
\newcommand{\tausop}{\psi_{S^\prime} (0)}
\newcommand{\Hst}{H_S  (t)}
\newcommand{\Hstp}{H_{S^\prime} \, (t)}
\newcommand{\pjss}{P_{j (S, S^\prime)}}
\newcommand{\ssp}{(S, S^\prime)}
\renewcommand{\mid}{\Big|}
\newcommand{\twomid}{\mid\mid}
\newcommand{\ket}[1]{|#1\rangle}
\newcommand{\bra}[1]{\langle #1|}
\newcommand{\braket}[2]{\langle #1|#2\rangle}
\newcommand{\fsqL}{\frac{1}{\sqrt{L}}}

\newcommand{\eilvvarp}{ e^{-i L \varphi}}

\newcommand{\eivvarp}{e^{-i  \varphi}}
\newcommand{\kpss}{\ket{\psi_1 (t)}}
\newcommand{\kpsss}{\ket{\psi_2 (t)}}
\newcommand{\kpsA}{\ket{\psi_A (t)}}
\newcommand{\kpsB}{\ket{\psi_B (t)}}
\newcommand{\kps}{\ket{\psi (t)}}
\newcommand{\fradvar}{\frac{d \varphi}{2\pi}}

\newcommand{\sulim}{\sum\limits}
\newcommand{\bvar}{|B (\varphi)|}
\newcommand{\avar}{|A (\varphi)|}
\newcommand{\rvee}{| R (E) |  }
\newcommand{\inlim}{\int\limits}
\newcommand{\ldrt}{d_r (t)}
\newcommand{\tpare}{T (E)}
\newcommand{\tparo}{T (0)}
\newcommand{\apara}{A (\varphi)}
\numberwithin{equation}{section}
\title{A Quantum Algorithm for the Hamiltonian NAND Tree}
\author{E. Farhi$^\ast$, J. Goldstone$^\ast$, S. Gutmann$^\dagger$\\
\it\small $^\ast$Center For Theoretical Physics\\
\it\small Massachusetts Institute of Technology\\
\it\small Cambridge, MA 02139\\[1ex]
\it\small $^\dagger$ Department of Mathematics\\[-.5ex]
\it\small  Northeastern University\\[-.5ex]
\it\small Boston, MA 02115
}
\date{}
\begin{document}
\maketitle
%\section{}
%\subsection{}

\begin{abstract}
We give a quantum algorithm for the binary NAND tree problem in the 
Hamiltonian oracle model.  The algorithm uses a continuous time 
quantum walk with a run time proportional to $\sqrt{N}$.  We also 
show a lower bound of $\sqrt{N}$ for the NAND tree problem in the 
Hamiltonian oracle model. \vtop{ \hsize= 3.5in \small 
\flushright{MIT-CTP/3813}}
%\flushright{MIT-CTP/3813}%\hspace{3in} MIT-CTP/3813
\end{abstract}

\section{Introduction}

\begin{figure}[ht]
\begin{center}
\includegraphics[]{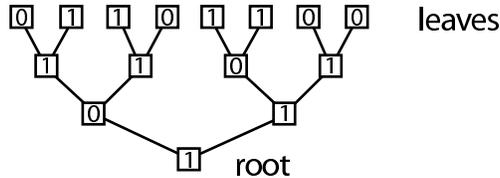}
\caption{A classical NAND tree.}
\label{Farfig1}
\end{center}
\end{figure}

The NAND trees in this paper are perfectly bifurcating trees with N 
leaves at the top and depth $n=\log_2(N)$.  Each leaf is assigned a 
value of 0 or 1 and the value of any other node is the NAND of the 
two connected nodes just above.  The goal is to evaluate the value at 
the root of the tree. An example is shown in figure (\ref{Farfig1}). 
Classically there is a randomized algorithm that succeeds after 
evaluating only (with high probability) $N^{.753}$ of the leaves. 
This algorithm is known to be best possible. See \cite{Farhi:Hoyer} 
and references there.

As far as we know, no quantum algorithm has been devised which 
improves on the classical query complexity.  However there is a 
quantum lower bound of $\sqrt{N}$ calls to a quantum oracle 
\cite{Farhi:Hoyer}.
In this paper we are not working in the usual quantum query model but 
rather with a Hamiltonian oracle \cite{Farhi:1996na, Farhi:Mochon} 
which encodes the NAND tree instance. We will present a quantum 
algorithm which evaluates the NAND tree in a running time 
proportional to $\sqrt{N}$. We also prove a lower bound of $\sqrt{N}$ 
on the running time for any quantum algorithm in the Hamiltonian 
oracle model.

\begin{figure}[ht]
%\begin{center}
\includegraphics[scale=.85]{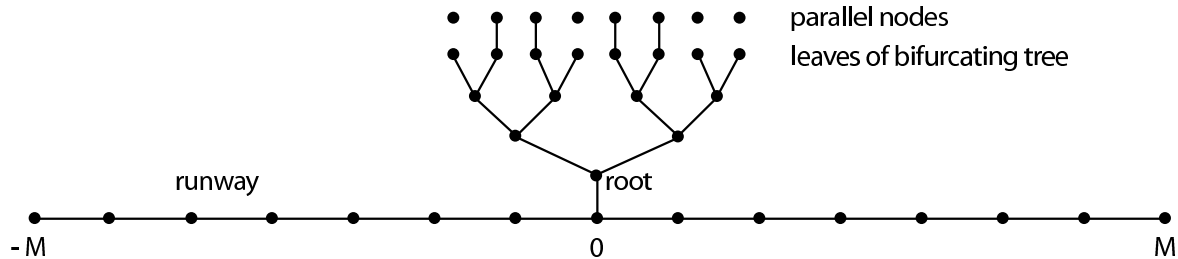}
\caption{The full Hamiltonian $H_O + H_D$}
\label{Farfig2}
%\end{center}
\end{figure}

Our quantum algorithm uses a continuous time quantum walk on a graph 
\cite{Farhi:1997jm}.  We start with a perfectly bifurcating tree of 
depth n and one additional node for each of the N leaves.  To specify 
the input we connect some of these N pairs of nodes. A connection 
corresponds to an input value of 1 on a leaf in the classical NAND 
tree and the absence of a connection corresponds to a 0. See the top 
of figure (\ref{Farfig2}).   Next we attach a long line of nodes to 
the root of the tree. We call this long line the ``runway".  See the 
bottom of figure (\ref{Farfig2}).  The Hamiltonian for the continuous 
time quantum walk we use here is minus the adjacency matrix of the 
graph. As usual with continuous time quantum walks,  nodes in the 
graph correspond to computational basis states.

  %\newpage

\begin{figure}[ht]
\begin{center}
\includegraphics[scale=.85]{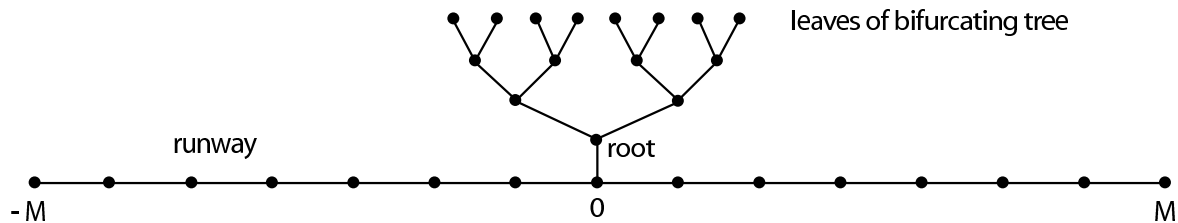}
\caption{The oracle independent driver Hamiltonian $H_D$}
\label{Farfig3}
\end{center}
\end{figure}

We can decompose this Hamiltonian into an oracle, $H_O$, which is 
instance dependent  and a driver, $H_D$, which is instance 
independent. $H_D$ is minus the adjacency matrix of the perfectly 
bifurcating tree of depth n whose root is attached to node 0 of the 
line of nodes running from $-M$ to $M$. We will take  $M $ to be very 
large. See figure (\ref{Farfig3}).

\begin{figure}[ht]
\begin{center}
\includegraphics[scale=.85]{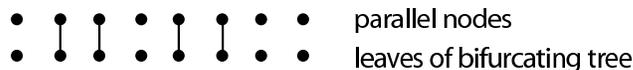}
\caption{The Hamiltonian oracle $H_O$}
\label{Farfig4}
\end{center}
\end{figure}

$H_O$ is minus the adjacency matrix of a graph consisting of the 
leaves of the bifurcating tree and the parallel set of N other nodes. 
Each leaf in the tree is connected or not to its corresponding node 
in the set above.  See figure (\ref{Farfig4}).   The quantum problem 
is: Given the Hamiltonian oracle $H_O$,  evaluate the NAND tree with 
the corresponding input.

Our quantum algorithm evolves with the full Hamiltonian $H_O + H_D$, 
which is  minus the adjacency matrix of the full graph illustrated in 
figure (\ref{Farfig2}).   The initial state is a carefully chosen 
right moving packet of length L localized totally on the left side of 
the runway with the right edge of the packet at node $0$. It will 
turn out that $L$ is of order $\sqrt{N}$.  We take M to be much 
larger than $L$, say of order $L^2$.  We now let the quantum system 
evolve and wait a time  $L/2$ which is the time it would take this 
packet to move a distance $L$ to the right if the tree were not 
present.  We then measure the projector onto the subspace 
corresponding to the right side of the runway.  If the quantum state 
is found on the right we evaluate the NAND tree to be 1 and if the 
quantum state is not on the right we evaluate the NAND tree to be 0.

We have chosen our right moving packet to be very narrowly peaked in 
energy around $E=0$. (Note that $E=0$ is not the ground state but is 
the middle of the spectrum.)  The narrowness of the packet in energy 
forces the packet to be long.  If we did not attach the bifurcating 
tree at node 0, the packet would just move to the right and we would 
find it on the right when we measure.  The algorithm works because 
with the tree attached the transmission coefficient at $E=0$ is 0 if 
the NAND tree evaluates to 0 and the transmission coefficient at 
$E=0$ is 1 if the NAND tree evaluates to 1.  The transmission 
coefficient is a rapidly changing function of $E$ but for $|E| < 1/ 
(16 \sqrt{N})$ the transmission coefficient is not far from its value 
at $E=0$. To guarantee that the packet consists mostly of energy 
eigenstates with their energies in this range, we take $L$ to be of 
order $\sqrt{N}$. This determines the $\sqrt{N}$ run time of the 
algorithm.

Our algorithm uses the driver Hamiltonian $H_D$ to evaluate the NAND 
tree.  An arbitrary algorithm can add any instance independent $H_A 
\, (t) $ to $H_O$ and work in the associated Hilbert space.  We will 
show that for any choice of instance independent $H_A \, (t)$ the 
running time required to evaluate the NAND tree associated with $H_O$ 
is of order at least $\sqrt{N}$.

\section{Motion on the Runway}
Here we describe the evolution of a quantum state initially localized 
on the left side of the runway in figure (\ref{Farfig2}) headed to 
the right.  M is so large that we can take it to be infinite as is 
justified by the fact that the speed of propagation is bounded. 
First consider the infinite runway  with integers $r$ labeling the 
sites. The tree is attached at $r=0$. We then have for all $r$ not 
equal to 0:
\begin{equation}%equation 1
H\, \ket{r} = -\ket{r+1} -\ket{r-1} \qquad r \ne 0 .
\label{farh07eq1}
\end{equation}
For $\theta > 0, e^{ir\theta}$ and $e^{-ir\theta}$ correspond to the 
same energy
%A state whose components on the runway are $e^{ir\theta}$ is a 
(nonnormalized) eigenstate of equation (\ref{farh07eq1}) with energy
\begin{equation}
E (\theta) = -2 \, \cos \, \theta
\label{farh07eq2}
\end{equation}
  but the first is a right moving wave and the second is a left moving 
wave.  We are interested in a packet, that is, a spatially finite 
superposition of energy eigenstates, which is incident from the left 
on the node 0 and the attached tree.  This packet will scatter back 
and also transmit to the right side of the runway.  The packet is 
dominated by energy eigenstates, $\ket{E}$, of the form on the runway
\begin{alignat}{5}%eq 3
\braket{r}{E} &= \left[e^{ir\theta} + R (E)\, e^{-ir\theta}\right] 
&\quad \mbox{for} \ r \leq 0, \label{farh07eq3}\\[1ex]
\braket{r}{E} &= T  (E) e^{ir\theta} &\quad \mbox{for}  \ r \geq 0 .
\label{farh07eq4}
\end{alignat}
(The states $\ket{E}$ do not vanish in the tree.)  There are other 
energy eigenstates, but we will not need them. Furthermore from 
standard scattering theory we have
\begin{equation}
\braket{E (\theta)}{E (\theta^\prime)} = 2 \pi \delta (\theta - 
\theta^\prime) .
\label{farh07eq5}
\end{equation}

Looking at $r=0$ we see that
\begin{equation}%eq 5
1 + R \, (E) = T \, (E) .
\label{farh07eq6}
\end{equation}
The transmission coefficient $T(E)$ is determined by the structure of 
the tree.  In particular let
\begin{equation}%eq6
y\, (E) = \frac{\braket{\mbox{root}}{E}}{\braket{r = 0}{E}}
\label{farh07eq7}
\end{equation}
  where ``root" is the node immediately above $r=0$ on the runway, see 
figure (\ref{Farfig2}).
Applying the Hamiltonian at $\ket{r=0}$ gives
\begin{equation}%eq7
H \ket{r = 0} = -\ket{r = -1} - \ket{r= 1} -\ket{\mbox{root}}
\label{farh07eq8}
\end{equation}
and taking the inner product with $\ket{E}$ gives
\begin{equation}%eq 8
T (E) = \frac{2\, i \, \sin \theta}{2\, i\, \sin \theta + y \, (E)} \ .
\label{farh07eq9}
\end{equation}

In the next section we will show how to calculate $y(E)$ and show 
that if the NAND tree evaluates to 1 then $y(0)=0 $ meaning that 
$T(0)=1$ and if the NAND tree evaluates to 0 then $y(0)= \infty$  and 
$T(0)=0$.   Unfortunately we cannot build a state with only E=0 since 
it would be infinitely long.
Instead we build a finite packet that is long enough so that it is 
effectively a superposition of the states $\ket{E}$ with $E$ close 
enough to $0$ that $T (E)$ is close to $T(0)$. We introduce two 
parameters $\varepsilon$ and $D$ for which
%\begin{equation}
%| T(E) - T(0) | < D |E| \quad \mbox{for}\quad |E| < \varepsilon \, .
%\label{farh07eq10}
%\end{equation}
\begin{align}
| T(E) - T(0) | < D |E| \quad \mbox{for}\quad |E| < \varepsilon \, .
\label{farh07eq10}
\end{align}
The parameters $\varepsilon$ and $D$ do depend on the size of the 
tree, but for the remainder of this section we only use $\varepsilon 
\ll 1$ and $\varepsilon D \ll 1$.

The initial state we consider, $ \ket{\psi (0)}$, is given on the runway as
\begin{eqnarray}
\braket{r}{\psi(0)}& =& \frac{1}{\sqrt{L}} \ e^{i r \pi/2} \qquad 
\mbox{for} \qquad - L + 1 \leq r \leq 0\nonumber\\[1ex]
&=& 0 \qquad\hspace{.58in} \mbox{for} \qquad r \leq - L \quad 
\mbox{and} \quad r > 0
\label{farh07eq11}
\end{eqnarray}
and vanishes in the tree. For this state
\begin{equation}
\bra{\psi (0)} H \ket{\psi (0)} = 0
\label{farh07eq12}
\end{equation}
and
\begin{equation}
\bra{\psi (0)} H^2 \ket{\psi (0)} = \frac{5}{L}
\label{farh07eq13}
\end{equation}
so the spread in energy about $0$ is of order $1/\sqrt{L}$. However, 
for our purpose we will see that this state is effectively more 
narrowly peaked in energy, and (\ref{farh07eq13}) is really an 
artifact of its sharp edge. In fact most of the probability in energy 
is contained in a peak around 0 of width $1/L$. Because we take 
$\pi/2$ and not $-\pi/2$ in (\ref{farh07eq11}) we have a right moving 
packet with energies near 0.

Evolving with the full Hamiltonian of the graph we have
\begin{equation}
\ket{\psi (t)} = e^{- i H t} \ket{\psi{(0)}}.
\label{farh07eq14}
\end{equation}
We decompose $\ket{\psi (t)}$ into two orthogonal parts,
\begin{equation}
\ket{\psi(t)} = \ket{\psi_1 (t)} + \ket{\psi_2 (t)}
\label{farh07eq15}
\end{equation}
where
\begin{equation}
\ket{\psi_1 (t)} = \int\limits^{\frac{\pi}{2} 
+\varepsilon}_{\frac{\pi}{2} - \varepsilon}\, \frac{d \theta}{2 \pi}\ 
e^{-it E (\theta)}\, \ket{E (\theta)}\, \braket{E (\theta)}{\psi (0)}
\label{farh07eq16}
\end{equation}
and $\ket{E (\theta)}$ is given in (\ref{farh07eq3}) and 
(\ref{farh07eq4}) with the normalization (\ref{farh07eq5}). The state 
$\ket{\psi_2 (t)}$ is a superposition of the other eigenstates of 
$H$, that is, states incident from the left with $0 < \theta < 
\frac{\pi}{2} - \varepsilon$ and $\frac{\pi}{2} + \varepsilon < 
\theta < \pi$, states incident from the right and bound states with 
$|E| >2$ and $\braket{r}{E} \to 0$ exponentially as $r \to \pm 
\infty$ on the runway. We do not need the details of $\ket{\psi_2 
(t)}$ since we will show in the appendix that the norm of 
$\ket{\psi_1 (t)}$ is close to $1$ so the norm of $\ket{\psi_2 (t)}$ 
is near 0. In other words, $\kpss$ is a very good approximation to 
$\kps$. To ensure this we need

\begin{equation}
L \gg \frac{1}{\varepsilon}  \ .
\label{farh07eq17}
\end{equation}
We would then expect that at late enough times we will see on the 
right a packet like the incident packet, but multiplied by $T (0)$ 
and moving to the right, with the group velocity $d E/d \theta$ 
evaluated at $\theta = \pi/2$ which is $2$. In the appendix we show, 
if $t$ is not too big,  that for   $r>0$
\begin{equation}
\braket{r}{\psi (t)} = T (0) \, \braket{r - 2t}{\psi (0)}
\label{farh07eq18}
\end{equation}
plus small corrections. (In the appendix we also make sense of this 
equation for $2t$ not an integer.) To ensure (\ref{farh07eq18}) we 
also need
\begin{equation}
L \gg D^2 \varepsilon .
%\label{farh07new}
\end{equation}
Since $\ket{\psi (0)}$ is a normalized state localized between $r= - 
L+1$ and $r = 0 $, (\ref{farh07eq18}) implies that for $t > L/2$
\begin{equation}
\sum\limits_{r>0} | \braket{r}{\psi (t)} |^2 = | T (0) |^2
\label{farh07eq19}
\end{equation}
plus small corrections so the cases $T (0) = 1$ and $T (0) = 0$ can 
be distinguished by a measurement on the right at $t > L/2$.

\section{Evaluating the transmission coefficient near E=0.}% section 3

 From the last section, equation (\ref{farh07eq9}), we see that we can 
find $T(E)$ if we know $y(E)$.  We now show how the structure of the 
tree recursively determines $y(E)$.  Consider the tree in figure 
(\ref{Farfig2}) ignoring the runway but including the runway node at 
$r=0$.   Except at the top or bottom of this tree every node is 
connected to two nodes above and one below.  See figure (5) where a, 
b, c and d are the amplitudes of $\ket{E}$ at the corresponding 
nodes.  Applying $H$ at the middle node yields:
\begin{equation}% eq 9
E \, a = -b -c -d
\label{farh07eq20}
\end{equation}
from which we get
\begin{equation}% eq 10
Y = -\frac{1}{E + Y^\prime + Y^{\prime\prime}}
\label{farh07eq21}
\end{equation}
%\begin{figure}[t]
%\begin{center}
%\includegraphics[]{fig5FINAL.eps}
%\caption{The amplitudes at 4 nodes in the middle of the tree in an energy eigenstate.}
%\label{Farfig5}
%\end{center}
%\end{figure}
\noindent where $Y=a/d, Y'=b/a$ and $Y''=c/a$. This is shown 
pictorially in figure (6). \\[2ex]
%\begin{figure}[ht]
%\begin{center}
%\quad\quad \includegraphics[]{fig6FINAL.eps}
%\caption{The recursion for $Y$.}
%\label{Farfig6}
%\end{center}
%\end{figure}
\centerline{\includegraphics[]{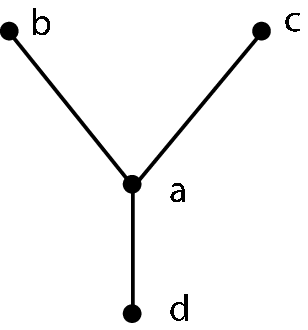}}\label{Farfig5}
\centerline{Figure 5: The amplitudes at 4 nodes in the middle of the 
tree in an energy eigenstate.}\\[2ex]
\centerline{\quad\quad \includegraphics[]{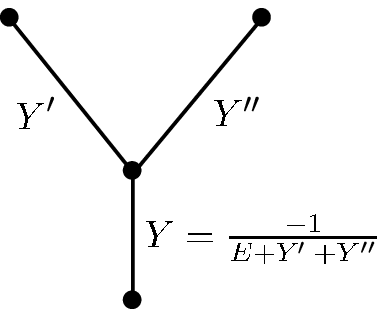}}
\centerline{Figure 6: The recursion for $Y$.}\\[3ex]
The $y$ we seek is $Y$ at the bottom of the tree.  To find it we 
recurse down from the top of the tree.  At the top of the tree there 
are 3 possibilities with the respective Y's obtained by applying the 
Hamiltonian.  The results are shown in figure (7). From figures (6) 
and (7) we see that $Y (-E) = - Y (E)$ so we can restrict attention 
to $E >0$.\\[1ex]
%\begin{figure}[ht]
%\begin{center}
%\includegraphics[scale=.85]{fig7FINAL.eps}
%\caption{The values of $Y$'s at the top of tree.}
%\label{Farfig7}
%\end{center}
%\end{figure}
\centerline{\includegraphics[scale=.85]{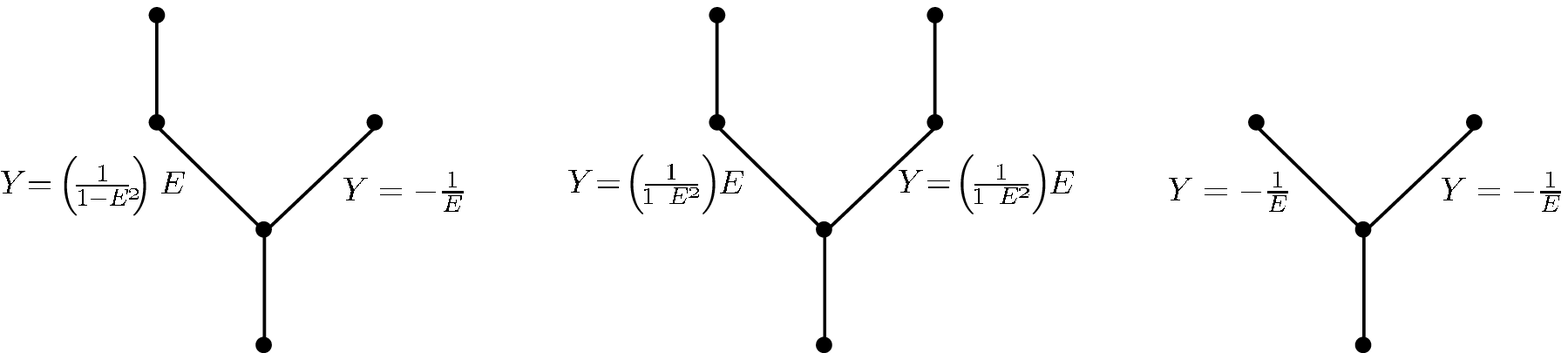}}
\centerline{Figure 7: The values of $Y$'s at the top of tree.}\\[1ex]

We now show that the recursive formula for computing the Y's at $E=0$ 
is in fact the NAND gate.   Looking at figure (7) and taking $E\to 0$ 
positively we get figure (8).  \\[3ex]
%\begin{figure}[ht]
%\begin{center}
%\includegraphics[scale=.85]{fig8FINAL.eps}
%\caption{The values of $Y$ at the top of the tree as $E\to 0^+$}
%\label{Farfig8}
%\end{center}
%\end{figure}
\centerline{\qquad\ \quad\ \ \includegraphics[scale=.85]{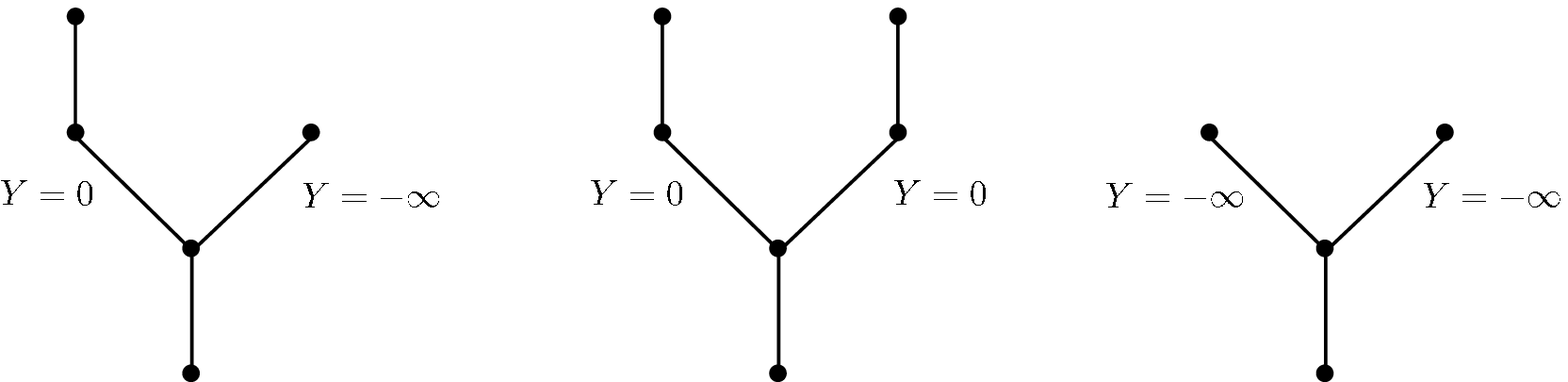}}
\centerline{\qquad\ Figure 8: The values of $Y$ at the top of the 
tree as $E\to 0^+$}\\[3ex]
We see that when we begin our recursion as $E\to  0+$  there are two 
initial values of Y which are -$\infty$ and 0.    Returning to figure 
(6) with the different possibilities for $Y' $and $Y^{\prime\prime}$ 
we get figure (9).  Identifying $Y=0$ with the logical value 1 and 
$Y=-\infty$ with the logical value 0 we see that figure (9) is a NAND 
gate.  Accordingly the value of $y(0)=Y(0)$ at the bottom of the tree 
(see \ref{farh07eq7}) in figure (\ref{Farfig2}) is the value of the 
NAND tree with the input specified at the top.\\[3ex]
%\begin{figure}[ht]
%\begin{center}
%\includegraphics[scale=.85]{fig9FINAL.eps}
%\caption{The recursion for Y at $E=0$ implements the NAND gate.}
%\label{Farfig9}
%\end{center}
%\end{figure}
\includegraphics[scale=.85]{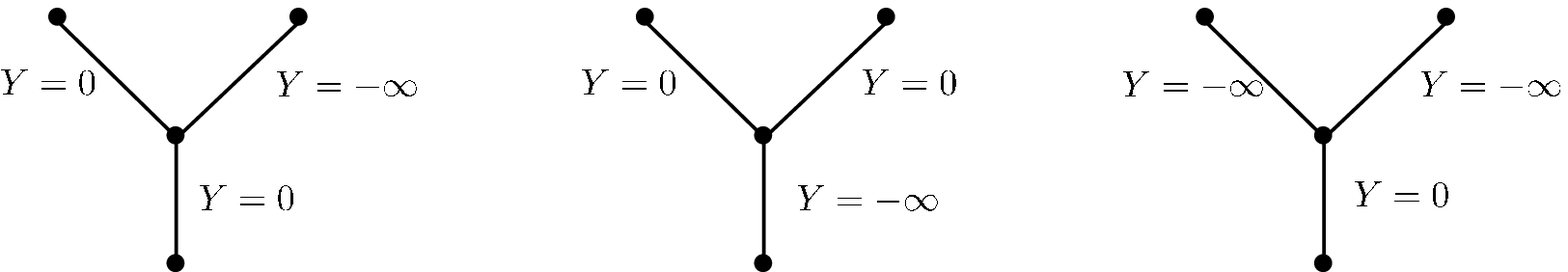}\\
\centerline{Figure 9: The recursion for Y at $E=0$ implements the 
NAND gate.}\\[2ex]

We now move away from $E=0$ to see how far from $E=0$ we can go and 
still have the value of the NAND tree encoded in $y(E)$ at the 
bottom.  Near $E=0$ it is convenient to write $Y(E)$ either as 
$a(E)\, E$ or as $ -1/(b(E) E)$ and we will find bounds on how large 
$a(E)$ and $b(E)$ can become as we move down the tree.   We observe 
that the coefficient of $E$ in $Y(E)$ (here we mean either $a(E)$ or 
$b(E))$ on any edge is an increasing function of the coefficients on 
the two edges above, as long as $|a  E|, |a^\prime E|, |b E|$ and 
$|b^\prime E|$ are not too big.   To see this use the recursive 
formula in figure (6). There are three cases and we obtain figure 
(10) which makes the increasingness clear. \\[3ex]
\includegraphics[scale=.85]{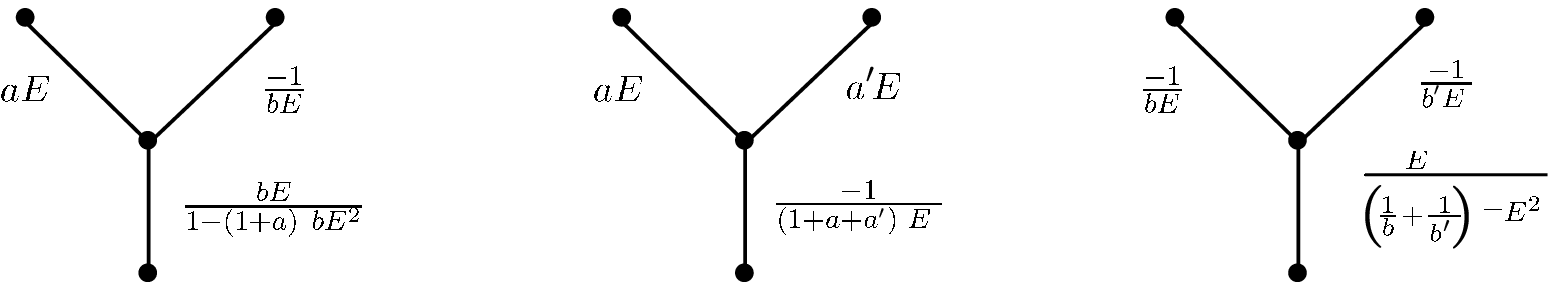}\\
\centerline{Figure 10: The coefficient of $E$ in $Y$ is an increasing 
function of $a, a^\prime, b \ \mbox{and}\ b^\prime$.}\\[3ex]

Now consider a piece of the bifurcating tree as depicted in figure 
(11). \\[3ex]
%\begin{figure}[ht]
%\begin{center}
%\includegraphics[]{fig11FINAL.eps}
%\caption{$Y$ determined by four $Y$'s above.}
%\label{Farfig11}
%\end{center}
%\end{figure}
\centerline{\includegraphics[]{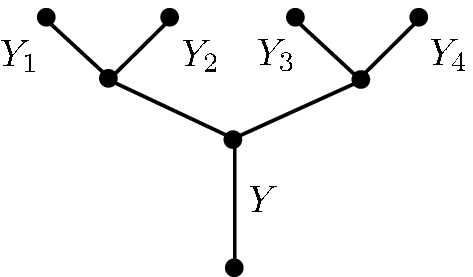}} \\
\centerline{Figure 11: $Y$ determined by four $Y$'s above.}\\[3ex]
There are 4 different Y's at the top of the piece, which determine  Y 
at the bottom of the piece.  Suppose for $0<E<2^{-n/2}/16$ each of 
the Y's at the top of the piece obeys
  \begin{alignat}{2}
0 \leq Y_i\, (E) \leq  c\, E \qquad & \mbox{or}&\qquad  0 \leq 
\frac{-1}{Y_i(E)} \leq\, c\, E
\label{farh07eq22}
\end{alignat}
for some positive $c$ with $c+1 \leq 4 \cdot 2^j$ where $j \leq n/2$. 
We will show
\begin{alignat}{2}
\qquad 0 \leq Y (E) \leq c^\ast \, E \qquad & \mbox{or}&\qquad 0 \leq 
\frac{-1}{Y (E)} \leq\, c^\ast\, E \hspace{.25in}
\label{farh07eq23}
\end{alignat}
where
\begin{equation}
(c^\ast + 1) = \frac{2(c+1)}{ 1- 2^{-n+2j}/4} .
\label{farh07eq24}
\end{equation}
There are six cases to consider and here we exhibit the          most 
dangerous case; see figure (12). \\[3ex]
\centerline{\includegraphics[]{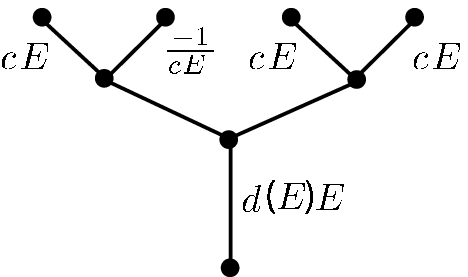}}
\centerline{Figure 12: The most dangerous case.} \\[3ex]
%\vspace{3ex}
The reason we can use the coefficient $c$ on the four top edges is 
the increasingness referred to earlier. Here $Y (E) = d (E) E$. 
Using the recursion from figure (10) we get
\begin{eqnarray}
d (E) &=& \frac{2 c + 1}{1- \left(1 + \frac{c}{1-(1+c) c E^2}\right) (2c+1) E^2
}\nonumber\\[1ex]
&\leq& \frac{2c + 1}{1-(1+ 2c) (2c + 1) E^2
}\label{farh07eq25}
\end{eqnarray}
%uisng $1 + c \leq 4 \cdot 2^{n/2}$ and $E^2 \leq \frac{1}{16^2} \, 2^n/n$. So
using $(1 + c) \leq 4 \cdot 2^{n/2}$ and $E^2 \leq \frac{1}{16^2} \, 
2^{- n}$. So
\begin{eqnarray}
d (E) & \leq&  \frac{2 c + 1}{1-4 (1+c)^2 E^2}\nonumber\\[1ex]
&\leq& \frac{2 (c +1)}{1-4 (1+ c)^2 E^2} - 1\nonumber \\[1ex]
&\leq& \frac{2 (c +1)}{1-2^{2j-n}/4} -1 \quad , \label{farh07eq26}
\end{eqnarray}
using $1+c \leq 4 \cdot 2^j$ and $E \leq 2^{-n/2}/16$. From 
(\ref{farh07eq26}) we get the left side of (\ref{farh07eq23}) with 
$c^\ast$ given in (\ref{farh07eq24}).  (From figure (10), the bounds 
on $c$ and $E$ also imply that the coefficients remain positive from 
one level to the next, so  $0 \leq Y (E)$.)
The other cases yield smaller coefficients and we will not run 
through them here.

At the top of the full tree $Y(E)=E/(1-E^2)$ or $Y(E)=-1/E$ as can be 
seen from figure (7). Given the restriction $O < E < 2^{-n/2}/16$ we 
can take
\begin{equation}
c_{\mbox{\scriptsize top}} = \frac{1}{1-2^{-n}} \, .
\end{equation}
Moving down the tree we get after $n/2$ iterations of
(\ref{farh07eq24}) with $j=1,2,\ldots n/2$,
\begin{equation}
\left( c^\ast_{\mathrm{bottom}}+1 \right) = \left( c_{\mathrm{top}} + 1
\right) \left( \frac{2}{1-2^{2-n}/4} \right) \left(
\frac{2}{1-2^{4-n}/4} \right) \ldots \left( \frac{2}{1-2^{n-n}/4}
   \right).
\end{equation}
Therefore
\begin{eqnarray}
c^\ast_{\mathrm{bottom}} &\leq& \frac{\left( 1+\frac{1}{1-2^{-n}} \right)
   2^{n/2}}{1-\frac{1}{4} \sum_{j=1}^{n/2} 2^{2j-n}} \\ \nonumber
& & \leq \frac{\left(1+\frac{1}{1-2^{-n}}\right)
   2^{n/2}}{1-\frac{1}{3}} \leq 4 \cdot 2^{n/2}
\end{eqnarray}
also justifying the assumption $c \leq 4 \cdot 2^j$ at each step.

%At the top of the full tree $Y(E)=E/(1-E^2)$ or $Y(E)=-1/E$ so $c$ 
%at the top can be taken to be $1/(1-2^{-n})$.  After $n/2$ iterations 
%we obtain the value of $c^\ast$ at the bottom of the tree:
%\begin{equation}
%c^\ast  \leq c^\ast  +1 = \left(1+\frac{1}{1-2^{-n}}\right) \ 
%2^{n/2} \, \frac{1}{\left(1- \frac{1}{2n}\right)^{n/2}}\ \leq 4 \cdot 
%2^{n/2} \ ,
%\label{farh07eq27}
%\end{equation}
%also justifying the assumption $c \leq 4 \cdot 2^{n/2}$ at each step.
%\begin{figure}[t]
%\begin{center}
%\includegraphics[]{fig12FINAL.eps}
%\caption{The most dangerous case.}
%\label{Farhfig12}
%\end{center}
%\end{figure}
This means that $|y(E)|$ at the bottom is either less than 
$|c^\ast_{\mbox{\scriptsize bottom}} E|$ or greater than 
$|1/(c^\ast_{\mbox{\scriptsize bottom}} E)|$ for $|E| < \ 
2^{-n/2}/16$. The crucial $\sqrt{N}$ arose because the coefficients 
of $Y (E)$ could barely more than double as we moved {\it two} levels 
down the tree.

Going back to the relation between $y(E)$ and the transmission 
coefficient given in (\ref{farh07eq9}) we summarize our results for 
this section:
\begin{center}
\begin{tabular}{|lccc|}
\hline\\
NAND = $0$ (reflect) & $|y| > \frac{1}{4\sqrt{N} |E|}$ & $|T| < 
8\sqrt{N} |E|$ & for  $|E| < \frac{1}{16\sqrt{N}}$\\[1.5ex]
%(reflect)&&\\
%\hline
NAND = 1  (transmit) & $|y| < 4 \sqrt{N} |E|$ & $|T-1| < 3\sqrt{N} 
|E|$ & for  $|E| < \frac{1}{16\sqrt{N}}$\\[2ex]
\hline
%(transmit)&&
\end{tabular}
\end{center}

\section{Putting it all together}
Here we combine the results of sections 2 and 3 and state the 
algorithm. First, the algorithm. Given the Hamiltonian oracle 
corresponding to an instance of the NAND tree problem we construct 
the full Hamiltonian $H_O +H_D$ which is minus the adjacency matrix 
of the graph in figure 2. We then build the initial state
\begin{equation}
\ket{\psi (0)} = \frac{1}{\sqrt{L}}\ \sum\limits^0_{r = - L +1}\, \ket{r}
\label{eq4-1}
\end{equation}
where $r$ is on the runway. We choose
\begin{equation}
L = \gamma \sqrt{N}
\label{eq4-2}
\end{equation}
with $\gamma$ an $N$ independent constant and $\gamma \gg 1$. We let 
the state evolve for a time
\begin{equation}
t_{\mbox{run}} = \frac{L}{2}
\label{eq4-3}
\end{equation}
and then measure the projector $P_+$ onto the right side of the runway
\begin{equation}
P_+ = \sum\limits^M_{r=1}\, \ket{r} \bra{r} .
\label{eq4-4}
\end{equation}
If the measurement yields $1$ we evaluate the NAND tree to be $1$ and 
if the measurement yields $0$ we evaluate the NAND tree to be $0$.

According to the results stated in Section 2, the probability of 
getting a measurement result of $1$ is very near $|T (0)|^2$ if
\begin{equation}
L \gg \frac{1}{\varepsilon} \quad \mbox{and} \quad    L \gg D^2 \varepsilon .
\label{eq4-5}
\end{equation}
 From the table at the end of section 3 we can take
\begin{equation}
\varepsilon = \frac{1}{16\sqrt{N}}
\label{eq4-6}
\end{equation}
and
\begin{equation}
D = 8 \sqrt{N}
\label{eq4-7}
\end{equation}
and then the choice $\gamma \gg 1$ ensures (\ref{eq4-5}).  From the 
end of the appendix we see that the error probability of the 
algorithm is $O \left(\frac{1}{\sqrt{L \varepsilon}}, \, D 
\sqrt{\frac{\varepsilon}{L}}, \, 
\left(\frac{\varepsilon}{L}\right)^{1/4}\right)$ which is $O 
\left(\sqrt{\frac{1}{\gamma}}\right)$ and independent of $N$. By 
choosing $\gamma$ large enough we can make the success probability as 
close to $1$ as desired.

\section{A lower bound for the Hamiltonian NAND tree problem via the 
Hamiltonian Parity problem}

Here we show that if the input to a NAND tree problem is given by the
Hamiltonian oracle $H_O$ described in section 1 then for an arbitrary
driver Hamiltonian $H_A(t)$, evolution using $H_O+H_A(t)$ cannot
evaluate the NAND tree in a time of order less than $\sqrt{N}$.  This
means that our algorithm which takes time $\sqrt{N}$ is optimal up to 
a constant.

In the usual query model, the parity problem with $\sqrt{N}$ 
variables can be embedded in a NAND tree with N leaves 
\cite{Farhi:Hoyer}.  To see this first consider 2 variables, a and b, 
and the 4 leaf NAND tree given in figure (13) which evaluates to $(1+ 
a +b)\!\!\mod{2}$. Using this we see that with 4 variables a,b,c,d 
the NAND tree in figure (14) evaluates to $(1 
+a+b+c+d)\!\!\!\mod{2}$. This clearly continues.   Since we know that 
the parity problem for $\sqrt{N}$ variables cannot be solved with 
less than of order $\sqrt{N}$ quantum queries, we know that the NAND 
tree problem cannot be solved with fewer than of order $\sqrt{N}$ 
quantum queries.\\[3ex]
\centerline{\includegraphics[]{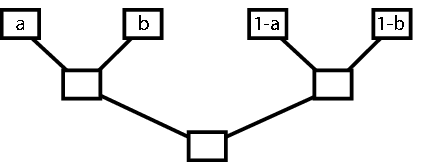}}
\centerline{Figure 13: This NAND tree evaluates to $(1 +a + b)\!\!\!\mod 2$.}
\vspace{3ex}

\centerline{\includegraphics{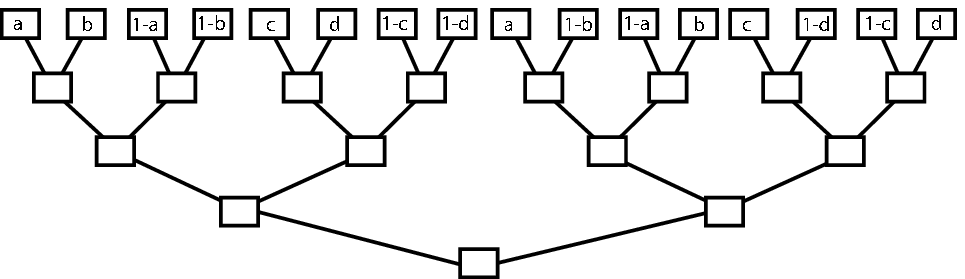}}
\centerline{Figure 14: This NAND Tree evaluates to $(1 + 
a+b+c+d)\!\!\!\mod 2$.}
\vspace{2ex}

In our Hamiltonian oracle model the input in figure (14) becomes the 
oracle depicted in figure (15).   Here the labels on the vertical 
edges on the top are always 0 or $-1$.  $A$ label $0$ means that the 
edge is not included $ \big($just as in figure (4)$\big)$ and 
accordingly the corresponding Hamiltonian matrix element is 0. $ A$ 
label $-1$   means that the edge is there (just as in figure (4)) and 
the corresponding matrix element of the Hamiltonian oracle is 
$-1$.\\[3ex]
\centerline{\includegraphics[scale=.75]{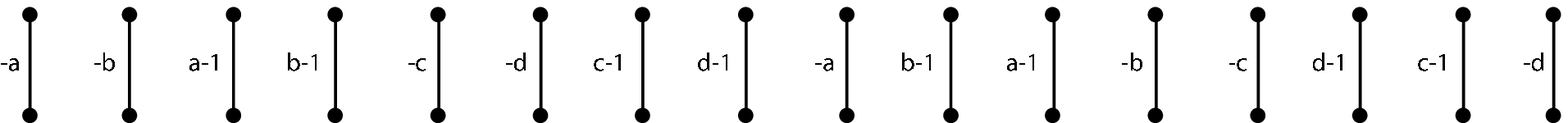}}\\[3ex]
\centerline{\parbox{6in}{Figure 15: The Hamiltonian Oracle for the 
NAND tree set up to evaluate $(1+a+b+c+d)\!\!\!\mod 2$. The 
coefficients are $0$ or $-1$ corresponding to the edge present or 
not.}}
\vspace{1ex}

We see that the Hamiltonian parity problem can be embedded in the 
Hamiltonian NAND tree problem.
We did this for the Hamiltonian NAND tree oracle considered in this 
paper, but it can also be done for the more general Hamiltonian NAND 
tree oracle described in the Conclusion.

We will now prove a lower bound for the Hamiltonian parity problem, 
in a general setting (see also \cite{Farhi:Mochon}), which can be 
used to obtain a $\sqrt{N}$ lower bound for the Hamiltonian NAND tree 
problem.
The oracle for the parity problem is a Hamiltonian of this form:
\begin{equation}
H_{O P}  = \sum\limits^K_{j= 1} \, H_j  .%\, (a_j)
\label{farh07eq28}
\end{equation}
The $H_j$  operate on orthogonal subspaces $V_j$ with
\begin{equation}
H_j = P_j \, H_j\, P_j
\label{farh07eq29}
\end{equation}
where $P_j$ is the projection onto $V_j$. Furthermore we assume $||H_j||\leq 
1$. For each $j$ there are two possible operators $H^{(a_j)}_j, a_j = 
0 \ \mbox{or}\ 1$. The string $a_1, a_2 \dots a_K$ is the input to 
the parity problem to be solved. In the Hamiltonian oracle model, an 
algorithm can use (\ref{farh07eq28}), but has no other access to the 
string $a$.
This is the most general form for the parity oracle Hamiltonian that 
we can imagine and it certainly includes the oracle we used to embed 
Hamiltonian parity in the Hamiltonian NAND tree.%\\[3ex]

Choose an arbitrary driver Hamiltonian $H_A(t)$.  Let S be an 
instance of parity, that is, a subset of ${1, 2,...K}$ with $a_j=1$ 
iff $j $ is an element of $S$.  Starting in an instance independent 
state $\ket{\psi(0)}$  we evolve for time T according to the 
Schrodinger equation
\begin{equation}
i\  \frac{d}{dt} \ket{\psi_S (t)} = \Hst \ket{\psi_S (t)}
\label{farh07eq30}
\end{equation}
using the total Hamiltonian $H_S(t)$
\begin{equation}
\Hst = g (t) \, H_{O  P} + H_A (t)
\label{farh07eq31}
\end{equation}
where $|g(t)|\leq 1$.  With the inclusion of the coefficient $g (t)$ 
it is clear that the Hamiltonian parity oracle model includes the 
quantum query parity model.

Let $\ket{\psi_S (T)}$ be the state reached at time $T$.  A 
successful algorithm for parity must have
\begin{equation}
\twomid \,\ket{\psi_S (T)} - \ket{\psi_{S^\prime}  (T)} \twomid \geq \delta
\label{farh07eq32}
\end{equation}
for some K independent $\delta > 0$ if the parity of $S$ and 
$S^\prime$ differ.  We now show that this suffices to force T to be 
of order K.  Our approach is an analogue to the analog analogue 
\cite{Farhi:1996na} of the BBBV method \cite{Bennett:1996iu}. We 
write $S \sim S^\prime$ if $S$ and $S^\prime$ differ by one element. 
Summing on $\ssp$ which differ by one element gives
\begin{eqnarray}
\frac{d}{dt}  \sns \twomid \,  \ket{\taus} - \ket{\tausp} \twomid^2 = 
\sns\, 2\, {\textrm I\textrm m}  \bra{\taus} \, \left(\Hst - 
\Hstp\right) \ket{\tausp} \nonumber\\[1ex]
=\sns\, 2\, {\textrm I\textrm  m} \, \bra{\taus}  \pm \triangle_{j 
\ssp} \ket{\tausp} \qquad\ \quad
\label{farh07eq33}
\end{eqnarray}
where $j\, \ssp$ is the element by which $S$ and $S^\prime$ differ, 
and $\triangle_j = g(t) \left(H_j^{(1)} - H_j^{(0)}\right)$. The $+$ 
sign means that $j (S, S^\prime)\, \epsilon\, S$ whereas the $-$ sign 
means that $j (S, S^\prime)\, \epsilon\, S^\prime$. Now we have
\begin{eqnarray}
\frac{d}{dt}  \sns \twomid \,  \ket{\taus} - \ket{\tausp} \twomid^2 
\leq 2\, \sns \mid \bra{\taus} \pjss\, \triangle_{j \ssp}\, \pjss 
\ket{\tausp} \mid\nonumber\\[1ex]
\leq 4 \sns \mid\mid \pjss \ket{\taus} \mid\mid \cdot \mid\mid \pjss 
\ket{\tausp} \mid\mid\
\label{farh07eq34}
\end{eqnarray}
since $|| \triangle_j || \leq 2$. Using $ a b \leq 
\left(\frac{1}{2}\, a^2 + \frac{1}{2}\, b^2\right)$ gives
\begin{eqnarray}
\frac{d}{dt}  \sns \twomid \,  \ket{\taus} - \ket{\tausp} \twomid^2 
\leq 2 \sns\ \left[ \twomid \pjss \ket{\taus} \twomid^2 + \twomid 
\pjss \ket{\tausp} \twomid^2\right]
\label{farh07eq35}
\end{eqnarray}
For fixed $S^\prime$, $j \ssp$ runs over $\{1, 2,\ldots K\}$ so
\begin{equation}
\sum\limits_{S \sim S^\prime, S^\prime \,  \mbox{fixed}} \twomid 
\pjss \ket{\tausp} \twomid^2 \quad \leq 1
\label{farh07eq36}
\end{equation}
and similarly for fixed $S$ and thus we have
\begin{equation}
\sns \ \left[\twomid \pjss \ket{\taus} \twomid^2 + \twomid \pjss 
\ket{\tausp} \twomid^2\right] \quad \leq 2 \cdot 2^K
\label{farh07eq37}
\end{equation}
since there $2^K$ possibilities for $S^\prime$ (or $S$).

We've shown that
\begin{equation}
\frac{d}{dt} \ \sns \twomid\ \ket{\taus} - \ket{\tausp} \twomid^2 
\quad \leq 4\cdot 2^K\ ,
\label{farh07eq38}
\end{equation}
and since
\begin{equation}
\sns \twomid\, \ket{\tauso} - \ket{\tausop}\twomid^2 \quad = 0
\label{farh07eq39}
\end{equation}
we can integrate to obtain
\begin{equation}
\sns \twomid \, \ket{\psi_S (T)} - \ket{\psi_{S^\prime}(T)} \twomid^2 
\leq \quad 4\cdot 2^K \cdot T\, .
\label{farh07eq40}
\end{equation}
For each $S^\prime$ there are $K$ choices of $S$, so a successful 
algorithm requires
\begin{equation}
\sns \twomid \, \ket{\psi_S (T)} - \ket{\psi_{S^\prime} (T)} 
\twomid^2 \quad \geq \quad 2^K \cdot K \cdot \delta
\label{farh07eq41}
\end{equation}
and we have the desired bound
\begin{equation}
T \geq K \, \delta/4.
\label{farh07eq42}
\end{equation}

\section*{Conclusion}
We are not working in the quantum query model but rather in the 
quantum Hamiltonian oracle model. In this model the programmer is 
given a Hamiltonian oracle of the form
\[
H_O = \sum\limits^{N}_{j=1} \, H_j
\]
where the $H_j$'s operate in orthogonal subspaces. Each $H_j$  is one 
of two possible operators $H_j^{\ (b_j)}$ with $b_j = 0$ or $1$ and 
the string $b_1\ldots b_N$ is the input to the classical NAND tree 
that is to be evaluated.
The quantum programmer is allowed to evolve states using any 
Hamiltonian of the form $g (t) \, H_O + H_A (t)$  where the 
coefficient $| g (t) | \leq 1$ and $H_A (t)$ is any instance 
independent Hamiltonian.
The programmer has no other access to the string $b$.

The algorithm presented in this paper uses a time independent $H_O +
H_D$ which is (minus) the adjacency matrix of a graph, so our
algorithm is a continuous time quantum walk.  We evaluate the NAND
tree in time of order $\sqrt{N}$ which is (up to a constant) the 
lower bound for this problem.

After the first version of this paper appeared, a $N^{1/2 + \epsilon}$
algorithm in the quantum query model was found\cite{Childs} using our
algorithm as a building block. As far as we can tell, our improvement
leaves that result unchanged.

\section*{Acknowledgement}
Two of the authors gratefully  acknowledge support from the National 
Security Agency (NSA) and the Disruptive Technology Office (DTO) 
under Army Research Office (ARO) contract W911NF-04-1-0216.

We also thank Richard Cleve for repeatedly  encouraging us to connect 
continuous time quantum walks with NAND trees, and for discussions 
about the Hamiltonian oracle model. We thank Andrew Landahl for 
earlier discussions of the NAND tree problem.

\appendix
\section{Appendix}
Here we flesh out the claims made in section 2 and put bounds on the 
corrections to the stated results. Since we work with $\theta$ close to 
$\pi/2$ it is convenient to write
\begin{equation}
\theta =   \varphi  + \pi/2
\label{eqapp1}
\end{equation}
and
\begin{equation}
E \, \left(\varphi + \pi/2\right) = 2 \sin \varphi
\label{eqapp2}
\end{equation}
so (\ref{farh07eq16}) becomes
\begin{equation}
\ket{\psi_1 (t)} = \int\limits^\varepsilon_{-\varepsilon} \, 
\frac{d\varphi}{2\pi} \ e^{-2i t \sin \varphi} \, \ket{E(\varphi + 
\pi/2)} \braket{E(\varphi + \pi/2)}{\psi (0)}
\label{eqapp3}
\end{equation}
where $\varepsilon$ is the small parameter introduced in 
(\ref{farh07eq10}). Using (\ref{farh07eq3}) and (\ref{farh07eq11}) 
and letting $r$ go to $-r$ gives
\begin{equation}
\braket{E(\varphi + \pi/2)}{\psi (0)} = \frac{1}{\sqrt{L}} \, 
\sum\limits^{L-1}_{r=0}\ \left\{e^{i r \varphi} + (-1)^r\, R^\ast 
(E)\, e^{- i r \varphi} \right\} .
\label{eqapp4}
\end{equation}
We now break $\ket{\psi_1 (t)}$ into two pieces
\begin{equation}
\ket{\psi_1 (t)} = \ket{\psi_A (t)} + \ket{\psi_B (t)}
\label{eqapp5}
\end{equation}
where
\begin{equation}
\ket{\psi_A (t)} = \int\limits^\varepsilon_{-\varepsilon} \, 
\frac{d\varphi}{2\pi}\ e^{-2 i t \sin \varphi} \, A (\varphi) \ket{E 
(\varphi + \pi/2)}
\label{eqapp6}
\end{equation}
with
\begin{equation}
A (\varphi) = \frac{1}{\sqrt{L}} \, \sum\limits^{L-1}_{r=0} \, e^{i 
r\varphi} = \frac{1}{\sqrt{L}}\ \  \frac{e^{i L \varphi} -1}{e^{i 
\varphi} -1}
\label{eqapp7}
\end{equation}
and
\begin{equation}
\ket{\psi_B (t)} = \inlim^\varepsilon_{-\varepsilon} \ 
\frac{d\varphi}{2\pi}\ e^{-2 i t \sin \varphi}\ R^\ast (E)\, B 
(\varphi)\ \ket{E (\varphi + \pi/2)}
\label{eqapp8}
\end{equation}
with
\begin{equation}
B (\varphi) = \frac{1}{\sqrt{L}}\ \sum\limits^{L-1}_{r=0} \ (-1)^r \, 
e^{-ir\varphi} = \fsqL \ \ \frac{1- (-1)^L \, \eilvvarp}{1 + 
\eivvarp} .
\label{eqapp9}
\end{equation}
Note that $\kpsA$ and $\kpsB$ are not orthogonal. Now
\begin{equation}
\bvar^2 < \frac{1}{L \cos^2 \frac{1}{2} \varepsilon} \quad \mbox{for} 
\quad |\varphi | < \varepsilon
\label{eqapp10}
\end{equation}
and $\rvee \leq 1$ so
\begin{equation}
\braket{\psi_B (t)}{\psi_B (t)} = \inlim^\varepsilon_{-\varepsilon} 
\, \fradvar \ \rvee^2\,  \bvar^2\,
\label{eqapp11}
\end{equation}
is of order $\varepsilon/L$ and we have
\begin{equation}
\twomid \kpsB \twomid = O \left(\sqrt{\frac{\varepsilon}{L}}\right) .
\label{eqapp12}
\end{equation}
On the other hand
\begin{equation}
\braket{\psi_A (t)}{\psi_A (t)} = \inlim^\varepsilon_{-\varepsilon} 
\fradvar \ \avar^2 .
\label{eqapp13}
\end{equation}
Now
\begin{equation}
\inlim^\pi_{- \pi} \, \fradvar \ \ \avar^2 = \frac{1}{L} \ 
\sulim^{L-1}_{r=0} \, 1 = 1
\label{eqapp14}
\end{equation}
while
\begin{equation}
\inlim^{-\varepsilon}_{-\pi} \, \fradvar\, \avar^2 + 
\inlim^{\pi}_\varepsilon\, \fradvar\ \avar^2 = \frac{1}{\pi} \ 
\inlim^\pi_\varepsilon \, d \varphi \, \frac{1}{L} \ \left(\frac{\sin 
\frac{1}{2} L\varphi}{\sin\frac{1}{2} \varphi}\right)^2  \ < 
\frac{\pi}{L \varepsilon}
\label{eqapp15}
\end{equation}
since $\sin \frac{\varphi}{2} > \frac{\varphi}{\pi}$ and $(\sin 
\frac{1}{2} L \varphi )^2 < 1$. Combining (\ref{eqapp13}), 
(\ref{eqapp14}) and (\ref{eqapp15}) gives
\begin{equation}
1 - \frac{\pi}{L \varepsilon} < \braket{\psi_A (t)}{\psi_A (t)} \leq 1
\label{eqapp16}
\end{equation}
so
\begin{equation}
\twomid \kpsA \twomid = 1 - O \left(\frac{1}{L \varepsilon}\right).
\label{eqapp17}
\end{equation}
We now require
\begin{equation}
L \gg \frac{1}{\varepsilon}
\label{eqapp18}
\end{equation}
so the norm of $\kpsA$ is close to 1.

Now
\begin{equation}
\twomid \kpss \twomid \geq \twomid \kpsA \  \twomid - \twomid \kpsB \twomid
\label{eqapp19}
\end{equation}
so
\begin{equation}
\twomid \kpss \twomid = 1 - O \left(\frac{1}{L \varepsilon} , 
\sqrt{\frac{\varepsilon}{L}}\right)
\label{eqapp20}
\end{equation}
where $O (a, b)$ means $O$ of the larger of $a$ and $b$. This 
justifies our claim in section 2 that $\kpss$ is a very good 
approximation to the true evolving state $\kps$. Since
\begin{equation}
\twomid \kpss \twomid^2 + \twomid \kpsss \twomid^2 = 1
\label{eqapp21}
\end{equation}
we also have
\begin{equation}
\twomid \kpsss \twomid = O \left(\frac{1}{\sqrt{L\varepsilon}} , 
\left(\frac{\varepsilon}{L}\right)^{1/4}\right).
\label{eqapp22}
\end{equation}

We measure the state $\kps$ on the right side of the runway, so we 
need a good estimate of
\begin{equation}
\bra{\psi (t)} P_+ \kps = \sulim_{r>0} \mid \braket{r}{\psi (t)} \mid^2 \, .
\label{eqapp23}
\end{equation}
Using
\begin{equation}
\kps = \kpsA + \kpsB + \kpsss
\label{eqapp24}
\end{equation}
and the bounds on the norms (\ref{eqapp12}), (\ref{eqapp17}) and 
(\ref{eqapp22})
we get
\begin{eqnarray}
\bra{\psi (t)} P_+ \kps = \bra{\psi_A (t))} P_+ \kpsA
  + O \left(\frac{1}{\sqrt{L\varepsilon}} , 
\left(\frac{\varepsilon}{L}\right)^{1/4}\right)
\label{eqapp26}
\end{eqnarray}
so we can use $\kpsA$ to estimate the probability of finding the 
state on the right at time $t$.

Now for $r\geq 0$ from (\ref{farh07eq4}) and (\ref{eqapp6})
\begin{equation}
\braket{r}{\psi_A (t)} = \inlim^{\varepsilon}_{-\varepsilon} \fradvar 
\ e^{-2 i t \sin \varphi}\, T (E) \, A(\varphi)\, i^r \, 
e^{ir\varphi} .
\label{eqapp27}
\end{equation}
Let
\begin{equation}
a_r (t) = T (0)\,  i^r \inlim^\pi_{-\pi} \fradvar \, e^{i (r -2t) 
\varphi}\, A (\varphi) .
\label{eqapp28}
\end{equation}
We want to show that $a_r (t)$ is a good approximation to 
$\braket{r}{\psi_A (t)}$. Let
\begin{equation}
\braket{r}{\psi_A (t)} = a_r (t) + b_r (t) + c_r (t)+ d_r (t)
\label{eqapp29}
\end{equation}
with
\begin{eqnarray}
b_r (t) &=& - i^r \,  T (0) \left\{\inlim^{-\varepsilon}_{-\pi}\, 
\fradvar + \inlim^\pi_\varepsilon \fradvar \right\} \ e^{i (r-2 t) 
\varphi} \, A (\varphi)\, , \label{eqapp30}\\[1ex]
  c_r (t) &=&  i^r \,  T (0) \inlim^{\varepsilon}_{-\varepsilon}\, 
\fradvar  (e^{-2 i t \sin \varphi} \, -e^{-2 i t \varphi}) \ e ^{i r 
\varphi}\, A (\varphi)
\label{eqapp31}
\end{eqnarray}
and
\begin{eqnarray}
\quad  d_r (t) &=&  i^r \,   \inlim^{\varepsilon}_{-\varepsilon}\, 
\fradvar\ \left(T(E) - T(0)\right)  \, e^{-2 i t \sin \varphi} e^{i r 
\varphi} \, A (\varphi).
\label{eqapp32}
\end{eqnarray}
We now show that $b_r (t), c_r (t)$ and $d_r (t)$ are small. First
\begin{equation}
\sulim^\infty_{r=0} |b_r (t)|^2 < \sulim^\infty_{-\infty}\, |b_r 
(t)|^2 =  |T(0)|^2 \cdot 2 \cdot \inlim^\pi_\epsilon \fradvar \, |A 
(\varphi)|^2
\label{eqapp33}
\end{equation}
by Parseval's Theorem. Using (\ref{eqapp15}) we have
\begin{equation}
\sulim^\infty_{r=0} |b_r (t)|^2 = O \left(\frac{1}{L\varepsilon}\right).
\label{eqapp34}
\end{equation}
Similarly
\begin{eqnarray}
\sulim^\infty_{r=0} |c_r (t)|^2 &<& |T(0)|^2 
\inlim^{\varepsilon}_{-\varepsilon}\, \fradvar  \mid e^{2i t (\varphi 
-\sin\varphi)} \, -1 \mid^2 |A (\varphi)|^2 \label{eqapp35}\\[1ex]
&= & | T(0) |^2  \inlim^{\varepsilon}_{-\varepsilon}\, \fradvar 
\left\{\frac{1}{9} \, t^2\, \varphi^6 + \cdot\cdot\cdot\right\}\ 
\frac{1}{L}\ \frac{\sin^2 \frac{1}{2}\, L \varphi}{\sin^2 
\frac{1}{2}\, \varphi}
\end{eqnarray}
However we take $t_{\mbox{\scriptsize run}}$ to be near $L/2$. As 
long as $L \varepsilon^3 \ll 1$ we have
\begin{equation}
\sulim^\infty_{r=0}  | c_r (t) |^2 = O (L \varepsilon^5) .
\label{eqapp35}
\end{equation}
To keep this small we need $L \varepsilon^5 \ll 1$, but this follows 
from our assumption that $L \varepsilon^3 \ll 1$ since $\varepsilon 
\ll 1$.  The assumption that $L \varepsilon^3 \ll 1$ helps us to 
establish the translation property in (2.18) which simplifies the 
picture of what is going on.
Also
\begin{equation}
\sulim^\infty_{r=0} \, | d_r (t) |^2 < 
\inlim^\varepsilon_{-\varepsilon}\, \fradvar\, | T (E) - T(0) |^2 | A 
(\varphi) |^2 .
\label{eqapp36}
\end{equation}

Using $|\tpare - \tparo | < D |E| \quad \mbox{for}\quad |E| < 
\varepsilon$ we get
\begin{eqnarray}
\sulim^\infty_{r=0} | \ldrt |^2 &<& D^2 
\inlim^{\varepsilon}_{-\varepsilon} \, \fradvar \, 4 \sin^2 \varphi 
\, \frac{1}{L} \ \frac{\sin^2 (\frac{1}{2} L \varphi)}{\sin^2 
(\frac{1}{2} \varphi)}\\[1ex]
&=& O (\frac{D^2 \varepsilon}{L})
\label{eqapp37}
\end{eqnarray}
from which we get our condition
\begin{equation}
L \gg D^2 \varepsilon .
\label{eqapp38}
\end{equation}

Now from (\ref{farh07eq11}) and (\ref{eqapp7}) we have
\begin{equation}
\braket{r}{\psi (0)} = i^r \, \inlim^\pi_{-\pi}\, \apara\, e^{i r 
\varphi}\, d \, \varphi
\label{eqapp39}
\end{equation}
So from (\ref{eqapp28}) we see that
\begin{equation}
a_r (t) = \tparo\ \braket{r-2t}{\psi (0)}
\label{eqapp41}
\end{equation}
but only when $2 t$ is an integer. However if $2 t = m + \tau$ with 
$m$ an integer and $0 \leq \tau < 1$,
\begin{eqnarray}
\sulim^{\infty}_{-\infty} | a_r (t) - a_r \left(\frac{m}{2}\right) 
|^2 &=& | T (0) |^2 \inlim^{\pi}_{-\pi} \, \fradvar\, \frac{1}{L}\, 
\frac{\sin^2 \frac{1}{2} L \varphi}{\sin^2 \frac{1}{2}  \varphi} \ | 
e^{i \tau \varphi} -1 |^2\nonumber\\
&=& O \left(\frac{1}{L}\right) \, .
\end{eqnarray}
We have shown that for $r > 0,  \braket{r}{\psi (t)}$ is well 
approximated by $ \braket{r}{\psi_1 (t)}$
which is well approximated by $\braket{r}{\psi_A (t)}$ which is well 
approximated by $a_r (t)$ which has the form (\ref{eqapp41}) so 
(\ref{farh07eq18}) is justified.

Furthermore, using (\ref{eqapp26}), (\ref{eqapp29}), (\ref{eqapp34}), 
(\ref{eqapp35}), (\ref{eqapp37}) and (\ref{eqapp41}) we have for $t > 
L/2$
\begin{equation}
\bra{\psi (t)}  P_+  \kps = |\tparo |^2 \ +  \ O 
\left(\frac{1}{\sqrt{L \varepsilon}} , \, D 
\sqrt{\frac{\varepsilon}{L}}, \, 
\left(\frac{\varepsilon}{L}\right)^{1/4}\right)
\end{equation}
which can be used to estimate the failure probability of the algorithm.

\end{document}